\documentclass[amsmath,amssymb,pre,showpacs]{revtex4}

\usepackage{graphicx}
\usepackage{dcolumn}
\usepackage{bm}
\usepackage{amssymb}
\usepackage{amsmath}
\usepackage{amsthm,amsfonts}

\begin{document}
\bibliographystyle{aip}

\title{Energy Rectification in Quantum Graded Spin Chains: Analysis of the XXZ Model}

\author{Lucas Schuab, Emmanuel Pereira}
 \email{emmanuel@fisica.ufmg.br}
\affiliation{Departamento de F\'{\i}sica--Instituto de Ci\^encias Exatas, Universidade Federal de Minas Gerais, CP 702,
30.161-970 Belo Horizonte MG, Brazil}
\author{Gabriel T. Landi}
\email{gtlandi@gmail.com}
\affiliation{Universidade Federal do ABC, 09210-580, Santo Andr\'e, Brazil}

\date{\today}

\begin{abstract}
In this work, with focus on the energy transport properties in quantum, low dimensional, graded materials, we address the investigation of the energy (and spin) current in XXZ open chains
with graded inner structures and driven out of equilibrium by magnetization pumping applied at the ends. We study several types of graded structures in different situations in order to show
a ubiquitous occurrence of energy rectification, even for the system under a homogeneous magnetic field. Due to technical difficulties, we carry out the computation for small chains, but we present arguments which indicate the extension of some results to larger systems.
Recalling the generic existence of energy rectification in classical, graded materials,
which are described by anharmonic chains of oscillators, and recalling also the anharmonicity of these XXZ models, which involve quartic terms in more transparent representation in terms of
fermionic creation and annihilation operators, we may say that our results extend the ubiquity of energy rectification occurrence in classical graded materials to the case of quantum systems.

\end{abstract}


\def \Z {\mathbb{Z}}
\def \R {\mathbb{R}}
\def \La {\Lambda}
\def \la {\lambda}
\def \ck {l}
\def \F {\mathcal{F}}
\def \M {\mathcal{M}}
\newcommand {\md} [1] {\mid\!#1\!\mid}
\newcommand {\be} {\begin{equation}}
\newcommand {\ee} {\end{equation}}
\newcommand {\ben} {\begin{equation*}}
\newcommand {\een} {\end{equation*}}
\newcommand {\bg} {\begin{gather}}
\newcommand {\eg} {\end{gather}}
\newcommand {\ba} {\begin{align}}
\newcommand {\ea} {\end{align}}
\newcommand {\tit} [1] {``#1''}


\maketitle

\let\a=\alpha \let\b=\beta \let\d=\delta \let\e=\varepsilon
\let\f=\varphi \let\g=\gamma \let\h=\eta    \let\k=\kappa \let\l=\lambda
\let\m=\mu \let\n=\nu \let\o=\omega    \let\p=\pi \let\ph=\varphi
\let\r=\rho \let\s=\sigma \let\t=\tau \let\th=\vartheta
\let\y=\upsilon \let\x=\xi \let\z=\zeta
\let\D=\Delta \let\G=\Gamma \let\L=\Lambda \let\Th=\Theta
\let\P=\Pi \let\Ps=\Psi \let\Si=\Sigma \let\X=\Xi
\let\Y=\Upsilon

\section{Introduction}

To understand the laws of transport starting from the underlying microscopic models is still a challenge in nonequilibrium statistical physics. In particular, the investigation
of energy or heat transport  is a fundamental problem of general interest and many open questions. It is intriguing to recall that Fourier's law, for example, a
keystone of the heat conduction theory, was proposed two hundred years ago, but the precise necessary and sufficient conditions for its validity are still
ignored \cite{BLR}. Such law is empirically observed to hold in general insulating three-dimensional systems: the seminal parabolic heat diffusion equation, present
in our basic text books, is  derived from it. However, exhaustive numerical simulations and also theoretical works \cite{LLP-D} show that Fourier's law fails for several
low dimensional systems ($d=1$ and $2$), in particular for models with total momentum conservation. Moreover, experimental works have confirmed the occurrence of this predicted anomalous
heat transport in nanomaterials related to some low dimensional models, such as carbon and boron-nitride nanotubes \cite{Chang2}, graphene  \cite{Gho}, nanowires \cite{Hoc}, etc.
In short,  the needfulness of detailed studies on the mechanism of transport in low dimensional structures is evident.

On the other side, in spite of these unanswered fundamental (and difficult) questions, the intense research in the area, together with the advance of nanotechnology, has opened the exciting
possibility to manipulate and control the energy current. Inspired by the amazing development  of modern electronics, thermal devices based on thermal diodes, e.g., thermal
transistors, gates, memories, etc, have been proposed \cite{BLiRMP} and experimentally built \cite{Chang1}. Again, open questions, such as the lacking of  efficient and feasible
diodes, stimulate concentrated research. In such direction, we recall the intense analysis of graded materials, which appear as genuine candidates for rectifiers \cite{WPC, CPC}.
Graded materials are systems in which some structure changes gradually in space: besides being abundant in nature, they can be manufactured and have attracted interest in
different areas, such as engineering, optics, material sciences, etc.

In the study of the microscopic mechanism of heat flow, it is also worth noting that most of the works  \cite{LLP-D, BLiRMP} involve classical dynamical
models. Consequently, this scarcity of quantum results, the present ambient of device miniaturization, together with the possibility of effects of quantum nature,
makes the detailed study of energy transport in genuine quantum models
a program of great importance. As an example of change in the transport properties due to the quantum/classical nature of the model, we recall that thermal rectification
has been observed in the quantum graded harmonic chain of oscillators with inner baths \cite{P1, Div1}, but it is absent in the classical version of the same model \cite{PL}. A natural
candidate for a quantum model describing these transport phenomena would be the quantum version of the chain of anharmonic oscillators, but even its classical version is already a problem of extreme difficulty
\cite{Kup}.

In the present work, with focus on the energy transport properties in quantum, low dimensional, graded materials,
we consider a more treatable genuine quantum model and address the investigation of the energy (as well as spin) current in  graded quantum XXZ open chains.
We investigate the one-dimensional graded XXZ model
driven out of equilibrium by the presence of pumping applied at the ends, precisely, by the coupling of the boundary spins of the chain to magnetization reservoirs. In short, as a first step, we study
the transport induced by a magnetization imbalance at the ends of the chain (it will  certainly be interesting, in a forthcoming work, to analyze the case of an energy imbalance, i.e., genuine thermal baths instead of magnetization reservoirs,  imposed at the boundaries).

We study different types of graded structures in the XXZ chain, and even different boundary conditions, in order to show that  energy rectification ubiquitously holds in these graded spin chains. It is worth
emphasizing that the conditions behind the onset of energy rectification are intricate: the simple existence of asymmetry in the systems by no means guarantees the occurrence of rectification.
We give a precise example.  Recall the classical harmonic chain of oscillators with inner stochastic reservoirs (which mimic the absent anharmonic potentials in more realistic models). This is an old model \cite{Bos},  which is, however,  recurrently studied \cite{BLL}. It has been rigorously proved that energy rectification is absent in any asymmetric version of this model (for example, with a graded mass distribution or with graded interparticle interactions) \cite{PLA}:  it seems to exist a preferential direction for the energy flow, but it does not.

We also need to stress that, besides the previously described
motivation related to quantum effects on the mechanism and properties of energy transport in low dimensional systems, there is a huge and recurrent interest in the detailed study of
the $1D$ XXZ chain by itself: it is an archetypal model to the investigation of open quantum systems with increasing attention in different areas, such as optics and cold-atoms, where it can
be experimentally realized \cite{N}. Many other fields of physics also involve problems related to open quantum systems, including nonequilibrium statistical physics, condensed matter,
quantum information, high-energy physics, etc \cite{BP}. We also need to emphasize that, in relation to specificities considered in the model (e.g., the graded structures and the boundary conditions),
the progress of nanotechnology and related experimental techniques allows us to manipulate different materials, including those with few elements, say, quantum bits. Even specific designs
of the coupling between systems and reservoirs are possible, and different many-body states and quantum phases may be prepared by taken proper quantum reservoirs \cite{Z}. In other words, even the study of specific
versions of the XXZ model is of considerable interest.

The rest of the paper is organized as follows. In section II, we introduce the model and describe the approach to be used in the computation of the currents in the steady state.
In section III, we derive some expressions for the currents and make some analysis by considering symmetry arguments. In section IV, we analyze the results due to the computations in the
the steady state. Section V is devoted to final remarks, and in the Appendix, we present some huge equations for the spin and energy currents under a homogeneous magnetic field.


\section{Model and Approach}

We consider here the one-dimensional quantum system given by the chain of $N$ particles with spin $\frac{1}{2}$  in the presence of an external magnetic field $B$, with nearest-neighbor
interaction given by the $XXZ$ model, namely, with Hamiltonian (for $\hbar = 1$)
\begin{eqnarray}
\mathcal{H} = && \sum_{i=1}^{N-1}\left\{ \alpha_{i,i+1}\left( \sigma_{i}^{x}\sigma_{i+1}^{x} + \sigma_{i}^{y}\sigma_{i+1}^{y} \right) + \Delta_{i,i+1}\sigma_{i}^{z}\sigma_{i+1}^{z} \right\}\nonumber \\
 && + \sum_{i=1}^{N} B_{i}\sigma_{i}^{z} ~, \label{hamiltonian}
\end{eqnarray}
where $\sigma_{i}^{\beta}$ ($\beta = x, y, z$) are the Pauli matrices and $B_{i}$ is the external magnetic field acting on site (particle) $i$.

To study the transport in the nonequilibrium steady state, we introduce a Markovian dynamics, as usual. We couple the chain to different magnetization baths, one at each end, so that the time evolution
of the system density matrix $\rho$ is given by a Lindblad quantum master equation \cite{BP}
\begin{equation}
\frac{d\rho}{d t} = i[\rho, \mathcal{H}] + \mathcal{L}(\rho) ~.\label{master}
\end{equation}
The dissipator $\mathcal{L}(\rho)$ describes the coupling with the baths: in the Lindblad form it is given by
\begin{eqnarray}
\mathcal{L}(\rho) &=& \mathcal{L}_{L}(\rho) + \mathcal{L}_{R}(\rho) ~, \nonumber\\
\mathcal{L}_{L,R}(\rho) &=& \sum_{s=\pm} L_{s}\rho L_{s}^{\dagger} -
\frac{1}{2}\left\{ L_{s}^{\dagger}L_{s} , \rho \right\} ~,\label{dissipator}
\end{eqnarray}
where, for $\mathcal{L}_{L}$, we have
\begin{equation}
L_{\pm} = \sqrt{\frac{\gamma}{2}(1 \pm f_{L})} \sigma_{1}^{\pm} ~\label{dissipator2},
\end{equation}
and a similar expression follows for $\mathcal{L}_{R}$, but with $\sigma_{N}^{\pm}$ and $f_{R}$ replacing  $\sigma_{1}^{\pm}$ and $f_{L}$. In the expressions above, $\{\cdot,\cdot\}$ denotes the anticommutator;
$\sigma_{j}^{\pm}$ are the spin creation and annihilation operators $\sigma_{j}^{\pm} = (\sigma_{j}^{x} \pm i\sigma_{j}^{y})/2$~; $\gamma$ is the coupling strength to the spin baths; $f_{L}$ and $f_{R}$ give the
driving strength: in the main cases considered here, we model the baths in terms of extra spins $\sigma_{0}^{z}$  and $\sigma_{N+1}^{z}$ linked to the chain, and, in such situation, $f_{L}$ and $f_{R}$ describe
different spin polarization at the boundaries: precisely, $f_{L} = \left<\sigma_{0}^{z}\right>$ and  $f_{R} = \left<\sigma_{N+1}^{z}\right>$. Written  in terms of $\sigma_{1}^{\pm}$ and $\sigma_{N}^{\pm}$,
the expression for the dissipator $\mathcal{L}(\rho)$ becomes
\begin{eqnarray}
\mathcal{L}(\rho) &=& \frac{\gamma}{4}\left\{(1+f_{L})\left[2\sigma_{1}^{+}\rho\sigma_{1}^{-} - \left( \sigma_{1}^{-}\sigma_{1}^{+}\rho  + \rho \sigma_{1}^{-}\sigma_{1}^{+}\right)\right]\right. \nonumber\\
                   &&   + (1-f_{L})\left[2\sigma_{1}^{-}\rho\sigma_{1}^{+} - \left( \sigma_{1}^{+}\sigma_{1}^{-}\rho  + \rho \sigma_{1}^{+}\sigma_{1}^{-}\right)\right]  \nonumber\\
 &&  + (1+f_{R})\left[2\sigma_{N}^{+}\rho\sigma_{N}^{-} - \left( \sigma_{N}^{-}\sigma_{N}^{+}\rho  + \rho \sigma_{N}^{-}\sigma_{N}^{+}\right)\right] \nonumber\\
                  &&  \left.  + (1-f_{R})\left[2\sigma_{N}^{-}\rho\sigma_{N}^{+} - \left( \sigma_{N}^{+}\sigma_{N}^{-}\rho  + \rho \sigma_{N}^{+}\sigma_{N}^{-}\right)\right] \right\}~.
\end{eqnarray}

One case with a different dissipator, namely, the model with twisted XY boundary gradients, is shortly examined in Sec.IV.

The derivation of the Lindblad master equation above is straightforward. For example, one may use the repeated interactions scheme as minutely described in appendices A and B of Ref.\cite{Landi}. The procedure
starts by taking an enlarged chain with two extra spins, labeled $0$ and $N+1$,  coupled to the boundaries of the original chain, i.e., to spins $1$ and $N$ respectively. The Hamiltonian for the
total enlarged system, as well as the time evolution, is determined after choosing some coupling interaction between the extra spins and the chain. It is assumed that in the initial time $t=0$ the baths (extra spins) are
decoupled from the chain,  so that the total density matrix factorizes as
$$
\rho_{T}(0) = \rho_{L} \rho(0) \rho_{R} ~,
$$
where $\rho_{L}$ and $\rho_{R}$ are the density matrices for the extra spins $0$ and $N+1$. The whole system (now with the extra spins) is allowed to evolve up to some time $\tau$. Then, the extra spins are
``discarded'' by taking a partial trace over $0$ and $N+1$, which leads to a new density matrix $\rho(\tau)$. New extra spins are taken from the baths and a new enlarged density matrix is built, as at time $t=0$,
with $\rho_{T}(\tau) = \rho_{L}\rho(\tau)\rho_{R}$. The process is repeated up to a time $2\tau$, and so on. Taking the relation between $\rho_{(n+1)\tau}$ and $\rho_{n\tau}$ obtained with the scheme and some further
manipulations, dividing  the difference $\left[\rho_{(n+1)\tau} - \rho_{n\tau}\right]$ by $\tau$, and taking the limit $\tau \rightarrow 0$, we obtain $d\rho/dt$ and the Lindblad master equation. See Ref.\cite{Landi} for a
complete and detailed derivation.

To obtain the steady state, i.e., the stationary density matrix $\rho_{S}$, reached as $t \rightarrow \infty$, we turn to the solution of Eq.(\ref{master}) with $d\rho/dt = 0$, i.e.,
\begin{equation}
0 = \frac{d\rho_{S}}{dt} = i\left[\rho_{S},\mathcal{H}\right] + \mathcal{L}(\rho_{S}) \equiv \mathcal{M} \rho_{S} ~.
\end{equation}
In other words, the nonequilibrium stationary state is given by the kernel (null space) of the linear operator $\mathcal{M}$.

Now, for convenience, we introduce the linear transformation $vec(A)$, the vectorization of a matrix $A$, which converts the matrix $A$ into a column vector. Precisely,
\begin{eqnarray*}
A = \left
[\begin{array}{cc} a & b \\
 c & d \end{array} \right]
~~\Rightarrow vec(A) = \left ( \begin{array}{c} a \\
c \\ b \\ d \end{array} \right ) ~.
\end{eqnarray*}
As well known, several properties follow for the vectorization, such as the compatibility with Kronecker products, namely,
$$
vec(ABC) = \left( C^{T}\otimes A\right)vec(B) .
$$

Next, we take the vectorization of the density matrix and related operators in the Lindblad master equation (\ref{master}).
From Eqs.(\ref{master}, \ref{dissipator}), it is clear that the R.H.S. of the Lindblad master equation is written in terms of products of matrices
$2^{N}\times 2^{N}$ such as $A\rho B$ (note that terms like $A\rho$ and $\rho B$ may be seen as $A\rho I$ and $I\rho B$, where $I$ is the identity matrix). Let us use the notation
$$
|\rho\rangle \equiv vec(\rho) ,
$$
for the vector $|\rho\rangle$ with $2^{2N}$ coordinates. Hence, the new notation for the master equation becomes
$$
\frac{d|\rho\rangle}{dt} = M |\rho\rangle ~,
$$
where, we stress, the linear operator $M$ acts on a vector space of dimension $2^{2N}$, i.e.,  it is a $2^{2N}\times 2^{2N}$ matrix. As already said, the stationary state $|\rho_{S}\rangle$ is given by
the eigenvector of $M$ with eigenvalue zero. For the particular Lindblad master equation considered here, general results \cite{EvansProsen} guarantee that the remaining eigenvalues of $M$ have negative real
parts, and so, as $t \rightarrow \infty$, our system reaches indeed the stationary state.

Within such an approach, we may (in principle) find exact results. However, the size of $M$ rapidly increases with $N$, making difficult even numerical computations in large systems, in particular for inhomogeneous
(graded) models. Recalling that one of our main motivations is the investigation of a possible extension to graded quantum systems of the ubiquitous occurrence of energy rectification found in general anharmonic
classical models with graded structures, we restrict the investigation to small chains, with $N$ up to $8$. It is important to mention that the study of small systems,
minimalistic mathematical prototypes or toy models may provide key information in physics. In a similar context, we recall, for example, that the existence of thermal rectification in the quantum graded chain of
harmonic oscillators with inner baths (in contrast with  absence in the classical model) has been first discovered in a chain with 3 sites only \cite{P1}, and then confirmed in larger chains by means of numerical computations
in a further work \cite{Div1}.

\section{Currents and Preliminary Results}

The expressions for the spin and energy currents follow from the dynamics, i.e., from the Lindblad master equation (\ref{master}) which gives the time evolution, and from continuity equations. For the magnetization flow, these
continuity equations are
\begin{eqnarray}
\frac{ d\langle \sigma_{1}\rangle}{dt} &=& \langle J_{L} \rangle - \langle J_{1} \rangle ~, \nonumber \\
\frac{ d\langle \sigma_{i}\rangle}{dt} &=& \langle J_{i-1} \rangle - \langle J_{i} \rangle ~,  ~ 1< i < N ~, \nonumber \\
\frac{ d\langle \sigma_{N}\rangle}{dt} &=& \langle J_{N-1} \rangle - \langle J_{R} \rangle ~.
\end{eqnarray}
Hence, as said, using the master equation, setting $f_{L}= f = -f_{R}$, and the continuity equations above, we obtain
\begin{eqnarray}
\langle J_{j} \rangle &=& 2\alpha \langle \sigma_{j}^{x} \sigma_{j+1}^{y} - \sigma_{j}^{y}\sigma_{j+1}^{x} \rangle ~, ~2\leq j \leq N-1 ~, \nonumber \\
                     &=& 4i\alpha \langle \sigma_{j}^{+} \sigma_{j+1}^{-} - \sigma_{j}^{-}\sigma_{j+1}^{+} \rangle ~,  \label{spinc} \\
\langle J_{L} \rangle &=& \gamma \left( f - \langle \sigma_{1}^{z} \rangle \right) ~,  \nonumber \\
\langle J_{R} \rangle &=& -\gamma \left( f + \langle \sigma_{N}^{z} \rangle \right) ~.
\end{eqnarray}

In the steady state we observe a homogeneous flow through the chain, namely,
\begin{equation}
\langle J_{1} \rangle_{S} = \langle J_{2} \rangle_{S} = \ldots = \langle J_{N} \rangle_{S} \equiv \langle J \rangle ~.
\end{equation}

To describe the energy current, as performed in Ref.\cite{Mendoza-A}, we first split the Hamiltonian (\ref{hamiltonian}) (with $\alpha_{i,i+1} = \alpha,~\forall i$) as
\begin{eqnarray}
H &=&  \sum_{i=1}^{N-1} \varepsilon_{i,i+1} = \sum_{i=1}^{N-1} h_{i,i+1} + b_{i,i+1} ~,  \\
h_{i,i+1} &=& \alpha \left( \sigma_{i}^{x} \sigma_{i+1}^{x} + \sigma_{i}^{y}\sigma_{i+1}^{y} \right) + \Delta_{i,i+1} \sigma_{i}^{z} \sigma_{i+1}^{z} ~, \nonumber\\
b_{i,i+1} &=& \frac{1}{2} \left[ B_{i}\sigma_{i}^{z}(1+\delta_{i,1}) + B_{i+1}\sigma_{i+1}^{z}(1+\delta_{i+1,N}) \right]~,\nonumber
\end{eqnarray}
i.e., we separate the part related to the XXZ interaction from the part associated with the external magnetic field. For the inner sites, $2\leq i\leq N-1$, taking again the time
evolution given by the master equation and the continuity equation
\begin{equation}
\frac{ d\langle \varepsilon_{i,i+1}\rangle}{dt} = \langle F_{i} \rangle - \langle F_{i+1} \rangle ~,
\end{equation}
we obtain, for the energy current,
\begin{equation}
\langle F_{j} \rangle = i\langle [\varepsilon_{j-1,j}, \varepsilon_{j,j+1}] \rangle ~, ~2\leq j \leq N-1 ~.
\end{equation}
For the boundaries, the corresponding contributions  (from the analysis of $d\langle \varepsilon_{i,i+1}\rangle/dt$ at the ends) are
\begin{eqnarray}
\langle F_{1} \rangle &=& Tr \left({\mathcal L}_{L}(\rho) \varepsilon_{1,2} \right) ~, \nonumber\\
\langle F_{N} \rangle &=& -Tr \left({\mathcal L}_{R}(\rho) \varepsilon_{N-1,N} \right) ~,
\end{eqnarray}
which describe the energy flow from the left reservoir to the chain, and from the chain to the right reservoir, respectively.

As said above about the Hamiltonian, it is also convenient to split the energy current as
\begin{equation}
\langle F_{i}\rangle = \langle F_{i}^{XXZ}\rangle + \langle F_{i}^{B}\rangle ~.
\end{equation}
For the XXZ contribution, we have (for $2\leq j\leq N-1$)
\begin{eqnarray}
\lefteqn{\langle F_{j}^{XXZ} \rangle = i\langle [h_{j-1,j}, h_{j,j+1}] \rangle = \ldots } \nonumber\\
&=& 2\alpha \langle \alpha \left( \sigma_{j-1}^{y}\sigma_{j}^{z} \sigma_{j+1}^{x} - \sigma_{j-1}^{x}\sigma_{j}^{z}\sigma_{j+1}^{y}\right) \nonumber\\
&& + \Delta_{j-1,j}\left( \sigma_{j-1}^{z}\sigma_{j}^{x} \sigma_{j+1}^{y} - \sigma_{j-1}^{z}\sigma_{j}^{y}\sigma_{j+1}^{x}\right) \nonumber\\
&& + \Delta_{j,j+1}\left( \sigma_{j-1}^{x}\sigma_{j}^{y} \sigma_{j+1}^{z} - \sigma_{j-1}^{y}\sigma_{j}^{x}\sigma_{j+1}^{z}\right)\rangle  ~.\label{Jxxz}
\end{eqnarray}
Consequently, again for $2\leq j\leq N-1$,
\begin{eqnarray}
\langle F_{j}^{B} \rangle &=& i\langle [\varepsilon_{j-1,j}, \varepsilon_{j,j+1}] - [h_{j-1,j}, h_{j,j+1}] \rangle = \ldots  \nonumber\\
 &=& \frac{1}{2} B_{j}\langle J_{j-1} + J_{j}\rangle ~,
 \end{eqnarray}
 where $J_{j}$ is the spin current, previously described.

 For the sites at the boundaries, we have
 \begin{widetext}
 \begin{eqnarray}
\langle F_{1}^{XXZ} \rangle &=& Tr \left({\mathcal L}_{L}(\rho) h_{1,2} \right)
    = -\frac{\gamma}{2} \left( \langle h_{1,2}\rangle + \Delta_{1,2}\langle \sigma_{1}^{z}\sigma_{2}^{z}\rangle \right) + \gamma f\Delta_{1,2}\langle \sigma_{2}^{z}\rangle ~,\nonumber\\
\langle F_{N}^{XXZ} \rangle &=& -Tr \left({\mathcal L}_{R}(\rho) h_{N-1,N} \right)
    = \frac{\gamma}{2} \left( \langle h_{N-1,N}\rangle + \Delta_{N-1,N}\langle \sigma_{N-1}^{z}\sigma_{N}^{z}\rangle \right) - \gamma f\Delta_{N-1,N}\langle \sigma_{N-1}^{z}\rangle ~;\\
\langle F_{1}^{B} \rangle &=& Tr \left({\mathcal L}_{L}(\rho) b_{1,2} \right)
    = \gamma B_{1} \left( f - \langle\sigma_{1}^{z}\rangle \right) ~,\nonumber\\
\langle F_{N}^{B} \rangle &=& -Tr \left({\mathcal L}_{R}(\rho) b_{N-1,N} \right)
    = -\gamma B_{N} \left( f + \langle\sigma_{N}^{z}\rangle \right) ~.
\end{eqnarray}
\end{widetext}

Note that, from the expressions given above, the energy current $\langle F_{i}\rangle$ is related to the spin current $\langle J_{i}\rangle$ as
\begin{equation}
\langle F_{i}\rangle = \langle F_{i}^{XXZ}\rangle + \frac{B_{i}}{2}\left(\langle J_{i-1}\rangle + \langle J_{i}\rangle \right) ~.\label{flows}
\end{equation}

In some cases, we can describe properties of the currents {\it a priori}, i.e., before the explicit computations with the steady density matrix. For example, in Ref.\cite{PopLi}, Popkov and Livi show that
arguments of symmetry in open spin chains with general assumptions about baths and interactions, may be used to prove the vanishing of energy and/or spin currents in some cases, despite the presence of
large boundary gradients. See also Ref.\cite{Mendoza-A} for similar considerations and other interesting qualitative arguments for the study of currents in the homogeneous (nongraded) XXZ chain.

Now, we follow such strategy to establish some properties of the currents (results which are confirmed later by direct algebraic computation), and also to make transparent the difference between the homogeneous and
the graded versions of the XXZ chain. For clearness and completeness, we repeat some arguments already presented in these references \cite{PopLi, Mendoza-A}.

To start with the arguments of symmetry, we first turn to the time evolution given by the Lindblad master equation (\ref{master}). Denoting the right-hand side of Eq.(\ref{master}) by $\mathcal{M}(\rho)$, we observe that if a
unitary transformation $U$ leaves  $\mathcal{M}(\rho)$ invariant, i.e., if $U \mathcal{M}(\rho)U^{\dagger} = \mathcal{M}(U\rho U^{\dagger})$, then $\tilde{\rho}(t) = U\rho(t) U^{\dagger}$ becomes a new solution of Eq.(\ref{master}). Consequently, for a system with a unique steady state $\rho_{S}$, it follows that $\rho_{S} = U\rho_{S} U^{\dagger}$. And so, for any physical observable $\mathcal{O}$ measured in the steady state, we have
$$
\langle \mathcal{O}\rangle \equiv Tr (\mathcal{O}\rho_{S}) = Tr (\mathcal{O} U\rho_{S}U^{\dagger}) = \langle U^{\dagger}\mathcal{O}U\rangle ~.
$$
If $\mathcal{O}$ also presents some symmetry under the action of $U$, for example, if $U^{\dagger}\mathcal{O} U = - \mathcal{O}$, then we have $\langle\mathcal{O}\rangle = \langle -\mathcal{O}\rangle$, and so,
$\langle\mathcal{O}\rangle = 0$. I.e., the average of $\mathcal{O}$ must vanish in the steady state.

Let us analyze the simpler case of the homogeneous XXZ chain (i.e., $\alpha_{i,i+1} \equiv \alpha$, $\Delta_{i,i+1} \equiv \Delta ,~ \forall i$) in the absence of the external magnetic field $B$. It is not difficult to see that the transformation
$U = \Omega^{\alpha}R$ leaves $\mathcal{M}(\rho)$ invariant, where
$$\Omega^{\alpha} = \sigma_{1}^{\alpha} \otimes \sigma_{2}^{\alpha} \otimes \ldots \otimes \sigma_{N}^{\alpha}~,$$
$\alpha = x, y$ and $R$ is the reflection operator
\begin{equation*}
R\left( A_{1} \otimes B_{2} \otimes \ldots \otimes Z_{N} \right) = \left(  Z_{1}\otimes \ldots \otimes B_{N-1} \otimes A_{N} \right)R ~,
\end{equation*}
(see Ref.\cite{PopLi} for more details). And so, $\rho_{S} = U\rho_{S} U^{\dagger} = \Omega^{\alpha}R \rho_{S} R\Omega^{\alpha}$ . It can be also verified that $F_{i}^{XXZ}$ changes sign under $U$, i.e., $U^{\dagger} F_{i}^{XXZ}
U = - F_{i}^{XXZ}$. Consequently, we have $\langle F_{i}^{XXZ}\rangle = 0$, and so, for the homogeneous XXZ chain, in the absence of external magnetic field, the
total energy current vanishes in the steady state: $\langle F \rangle = 0$ (see Eq.(\ref{flows})).

If we introduce a homogeneous magnetic field in the chain, $B_{i} = B ,~ \forall i$, as explained in Ref.\cite{Mendoza-A}, the spin current (see Eq.(\ref{spinc}))
$$
\langle J_{j} \rangle = 4i\alpha \langle \sigma_{j}^{+} \sigma_{j+1}^{-} - \sigma_{j}^{-}\sigma_{j+1}^{+} \rangle ~,
$$
which is given by processes conserving the number of spin excitations, is not affected.

Similarly, $F_{j}^{XXZ}$ is given by terms such as (see Eq.(\ref{Jxxz}))
\begin{equation*}
\sigma_{j-1}^{z}\left( \sigma_{j}^{x} \sigma_{j+1}^{y} - \sigma_{j}^{y}\sigma_{j+1}^{x}\right) = 2i\sigma_{j-1}^{z}\left( \sigma_{j}^{+} \sigma_{j+1}^{-} - \sigma_{j}^{-}\sigma_{j+1}^{+}\right) ~,
\end{equation*}
and so, it is also independent of the external magnetic field. In conclusion, for the homogeneous XXZ chain in the presence of a homogeneous magnetic field $B$, as $\langle F_{j}^{XXZ}\rangle = 0$, we have
\begin{equation}
\langle F \rangle = B \langle J \rangle ~,
\end{equation}
that is, the total energy current is ruled by the spin current.

However, with the introduction of a graded structure in the XXZ, i.e., turning to the graded XXZ model, the picture changes. For example, in the case of a system in which $\Delta_{i,i+1}$ increases with $i$,
it is evident that the reflection $R$ completely modifies the Hamiltonian $\mathcal{H}$ (\ref{hamiltonian}), and so,  $\mathcal{M}(\rho)$ loses its invariance under the transformation $U=\Omega^{\alpha}R$. That is, the
symmetry property $\rho_{S} = U\rho_{S}U^{\dagger}$ is lost, and we cannot say that $\langle F^{XXZ}\rangle$ vanishes in the presence of a homogeneous magnetic field $B$ anymore. In fact, this imbalance of energy
in the interaction due to the increasing of $\Delta_{i,i+1}$ with $i$ indicates a preferential direction for the energy flow, and, even for $B=0$, $F^{XXZ}$ is not expected to vanish. Moreover, for $B=0$ the effect of
the asymmetry in $\Delta_{i,i+1}$ does not change if we invert the signs of all $S^{z}$, i.e.,
the direction of increasing energy interaction is not affected (as
well known, $S^{z} = \sigma^{z}/2$, for $\hbar = 1$).
Thus, in a graded chain in which the driving strength is $f$ for the left end and $-f$ for the right one, or in the same graded chain
with inverted baths (i.e, with $-f$ at left end and $f$ at right one), the energy flow and its direction will be the same. In the presence of a magnetic field $B$,  this picture is spoiled, but, anyway, we have the occurrence of
energy rectification, i.e., a preferential direction for the current. Let us make transparent this very important point. Note that, from Eq.(\ref{flows}), in the steady state and in the presence of a homogeneous magnetic
field we have $\langle F \rangle = \langle F^{XXZ} \rangle + B \langle J \rangle$, with  $\langle F^{XXZ} \rangle$ an even function of $f$, as argued above, and $\langle J \rangle$ an odd function of $f$
(the spin current direction is determined by the direction of the magnetization imbalance). Hence, the energy rectification (change in the
magnitude of $\langle F \rangle$ as we invert the baths) is clear: if we invert the sign of $f$, only one term of $\langle F \rangle$ changes the sign.

As a further comment, it is worth stressing that the symmetry arguments above (dependence on $f$) are not related to the size of the chain, and so, they indicate the occurrence of rectification in larger chains once proved its
existence in small systems, which is minutely described ahead.

In the next section, we confirm this scenario predicted by symmetry considerations and present more new information with the description of the currents computed from the steady density matrix.

\section{Steady State Computation: Solutions and Properties}

Before describing our findings from the computation work, it is wort recalling some recent results about rectification (but of the spin current) in homogeneous XXZ chains. In Ref.\cite{Landi}, the authors show the existence of spin current
rectification in the homogeneous XXZ chain (with $\alpha_{i,i+1} = \alpha$ and $\Delta_{i,i+1} = \Delta ,  \forall i$) under an inhomogeneous, linearly graded, magnetic field for small chains, up to $N=7$. Moreover, they show the vanishing of
rectification as $\Delta \rightarrow 0$, i.e., for the XX chain, and emphasize the correspondence to the well known result of absence of (energy) rectification in classical harmonic chains: written in terms of fermionic
creation and annihilation operators, the XX model contains only quadratic terms, whereas the XXZ model involves also a quartic term which is proportional to $\Delta$.

Turning to our results, we first describe the solutions for the graded XXZ chain under a homogeneous magnetic field. Given the huge algebraic computation, we start with a small chain, with $N=3$ and with the dissipator
targeting values of the $z$-spin at the boundaries - see Eqs.(\ref{hamiltonian}, \ref{dissipator2}). We take $\Delta_{1,2} = \Delta - \delta$ and $\Delta_{2,3} = \Delta + \delta$. The final expressions for the currents in
the steady state are presented in the Appendix. Analyzing the formulas, we note that, as expected, we may split the expression for the energy current as $\langle F \rangle = \langle F^{XXZ} \rangle + B \langle J \rangle$. For
$B=0$, we have that $\langle F \rangle$ becomes an even function of $f$, the driving strength, and it vanishes as $\delta \rightarrow 0$, i.e., without the graded asymmetry. On the other side, still for $B=0$,
the energy flow does not vanish in the graded chain ($\delta\neq 0$), and its direction is determined by the graded asymmetry: it does not depend on the direction of the driving strength, which, however, gives the direction of the spin current.

With the introduction of a homogeneous magnetic field $B$, as in the non-graded $\Delta$ case, the spin current does not change, whereas the energy current is modified. Again, as previously predicted and discussed,
there is rectification for the energy, but not for
the spin current.

In the present paper, we define the rectification
factor essentially as the sum of the energy flow in a given situation and the related flow as we invert the baths, divided by the difference of the flows. For convenience, we multiply the expression by 100, precisely,
\begin{equation}
\mathcal{R}_{E} \equiv 100 \times \left|\frac{\langle F \rangle + \langle F \rangle_{I}}{\langle F \rangle - \langle F \rangle_{I}}\right| ~.
\end{equation}
Such a definition is similar to those recurrently used in the study of rectification in quantum spin chains \cite{Landi}. In the case of a perfect diode, i.e., if we have some flow in a given direction but no flow in the chain with inverted baths,  we obtain $\mathcal{R}_{E} = 100$; in the case of equal flows, i.e., in the situation in which the inversion of the baths does not change anything, not even the direction of the flow, as is the case above with
$B=0$, we have $\mathcal{R}_{E} = \infty$.

As described by the formulas in the Appendix, the energy current, as well as the magnetization flow, depends on several parameters ($f, \alpha, \Delta, \delta, B$), which leads to a very varied scenario. As illustration,
we plot some curves to visualize the behavior of the energy rectification below, considering different parameters and/or regions.
See Fig.1 for the energy rectification versus the driving strength $f$ for different values of $\Delta$. We recall that for large, homogeneous chains, the cases $\Delta >1$ and $\Delta <1$ ($\alpha =1$) describe different situations: in the absence of a
magnetic field, the ground state diagram of this model engenders a gapped phase for $\Delta >1$, and a gapless one for $\Delta <1$, with consequences on the transport properties \cite{Mendoza-A}.
In Fig.2 we plot the energy rectification versus the interaction asymmetry $\delta$, for different values of external magnetic field $B$.

\begin{figure}[h]
\includegraphics[width=0.40\textwidth]{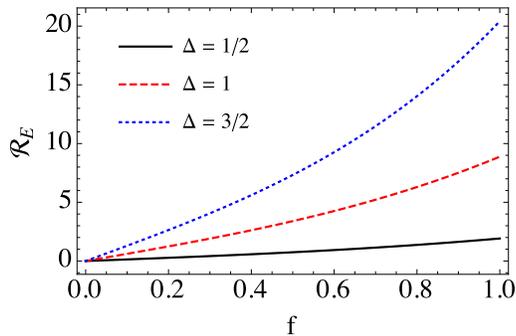}
\caption{(Color online) Energy rectification versus driving strength, for $N=3$ and graded interaction between $S_{i}^{z}$ and $S_{i+1}^{z}$. Here, $B=0.1$, $\alpha = 1$,  $\delta = 1/4$.}
\end{figure}

\begin{figure}[h]
\includegraphics[width=0.40\textwidth]{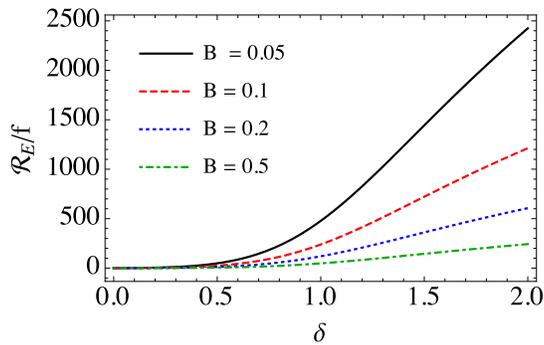}
\caption{(Color online) Energy rectification versus parameter of asymmetry, for $N=3$ and graded interaction between $S_{i}^{z}$ and $S_{i+1}^{z}$. Here, $\alpha = 1$,  $\Delta =0$, $f = 0.01$.}
\end{figure}

In order to make clear that the energy rectification remains for different system sizes, we describe below  the rectification factor  in the cases $N=3, 4, \ldots, 8$, for $\Delta =1/2$ and $\Delta =3/2$, and different
asymmetry parameter  $\delta$. See Fig.3. As already said in Sec.II, in a system with $N$ spins, the computation of the steady state
distribution involves vectors with $2^{2N}$ coordinates, which makes difficult numerical calculations for large $N$. To perform the numerical computations, we take $\alpha = 1$, $B = 0.1$, $f = 1$ and a) $\Delta = 1/2$, b) $\Delta = 3/2$.

\begin{figure}[h]
\includegraphics[width=0.40\textwidth]{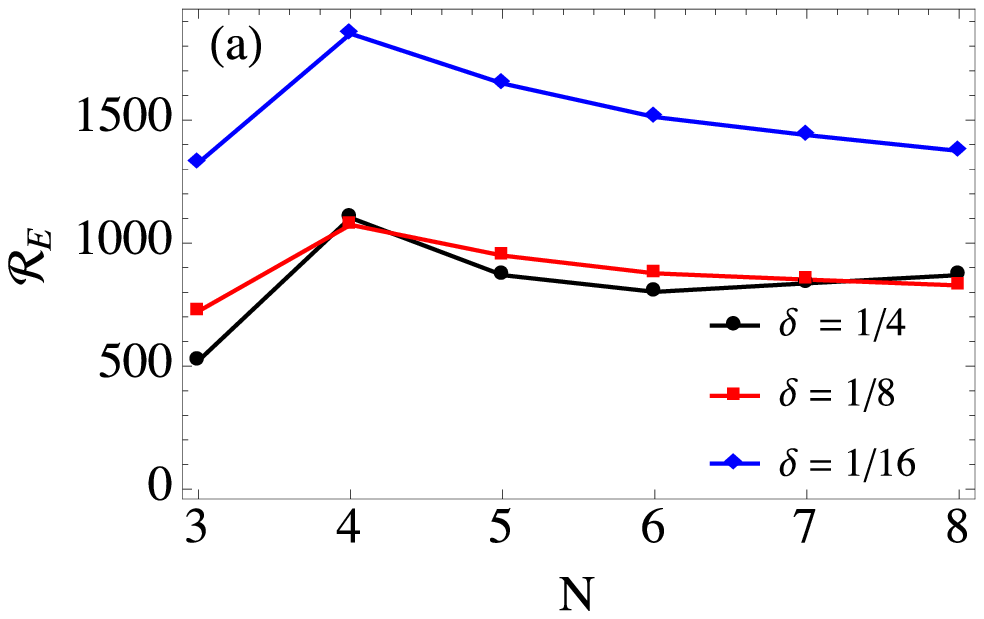}\quad
\includegraphics[width=0.40\textwidth]{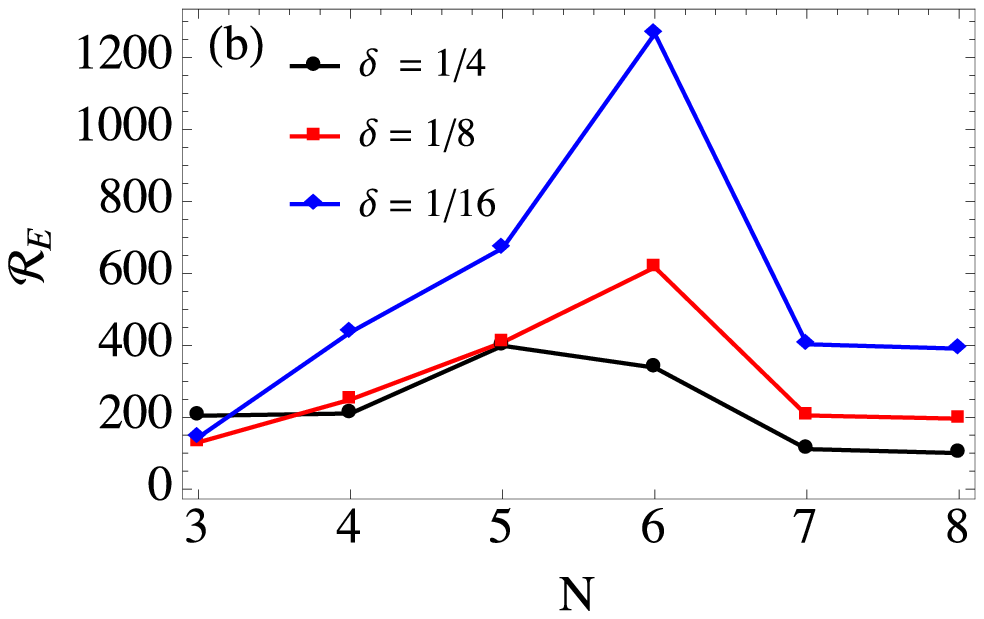}
\caption{(Color online) Energy rectification versus system size. Here,  $\alpha = 1$,  $B=0.1$, $f = 1$; and a) $\Delta = 1/2$, b) $\Delta= 3/2$.}
\end{figure}

In order to verify the robustness of the energy rectification due to intrinsic graded structures in the Hamiltonian interaction, we investigate the currents in cases of different graded interactions, but still
in the presence of a homogeneous magnetic field $B$. Precisely, for $N=3$, we study the following cases. a) Graded interaction between $S_{i}^{x}$ and $S_{i+1}^{x}$, $S_{i}^{y}$ and $S_{i+1}^{y}$ - i.e.,
$\alpha_{1,2} = \Lambda - \delta$,  $\alpha_{2,3} = \Lambda + \delta$; and homogeneous interaction between $S_{i}^{z}$ and $S_{i+1}^{z}$, i.e., $\Delta_{i,i+1} = \Delta$. b) Graded XXX model, i.e.,
 $\alpha_{1,2} = \Delta_{1,2} = \Delta - \delta$,
$\alpha_{2,3} = \Delta_{2,3} = \Delta + \delta$. c) Completely graded XXZ model, i.e., $\alpha_{1,2}  = 1 - \delta$, $\alpha_{2,3}  = 1 + \delta$,  $\Delta_{1,2} = \Delta - \delta$, $\Delta_{2,3} = \Delta + \delta$. For all the
cases, the results scenario is repeated, namely, there is no rectification for the spin current, but the energy current rectifies. See Fig.4 a), b) and c). Moreover, for all these cases in the particular situation of $B=0$, we still have the
same energy current as we invert the baths at the ends (changing $f \leftrightarrow -f$): precisely, the direction of the energy current does not invert with the inversion of the baths.

Besides these cases, we also study the previous $\Delta$-graded XXZ chain with $N=3$, but with different boundary conditions: precisely, we consider the model with twisted XY boundary gradients \cite{Pop2}, i.e. with
$\langle S_{0}^{x} \rangle = \kappa = \langle S_{N+1}^{y} \rangle$.
In the presence of a homogenous magnetic field, we observe again the rectification of the energy. See Fig.4 d).
Moreover, now it is also noted the onset of a spin current rectification.
As the parameter of asymmetry $\delta$ (responsible for the graded structure) and the magnetic field go to zero, the energy current vanishes, as expected \cite{Pop2}.


\begin{figure}[h]
\includegraphics[width=0.40\textwidth]{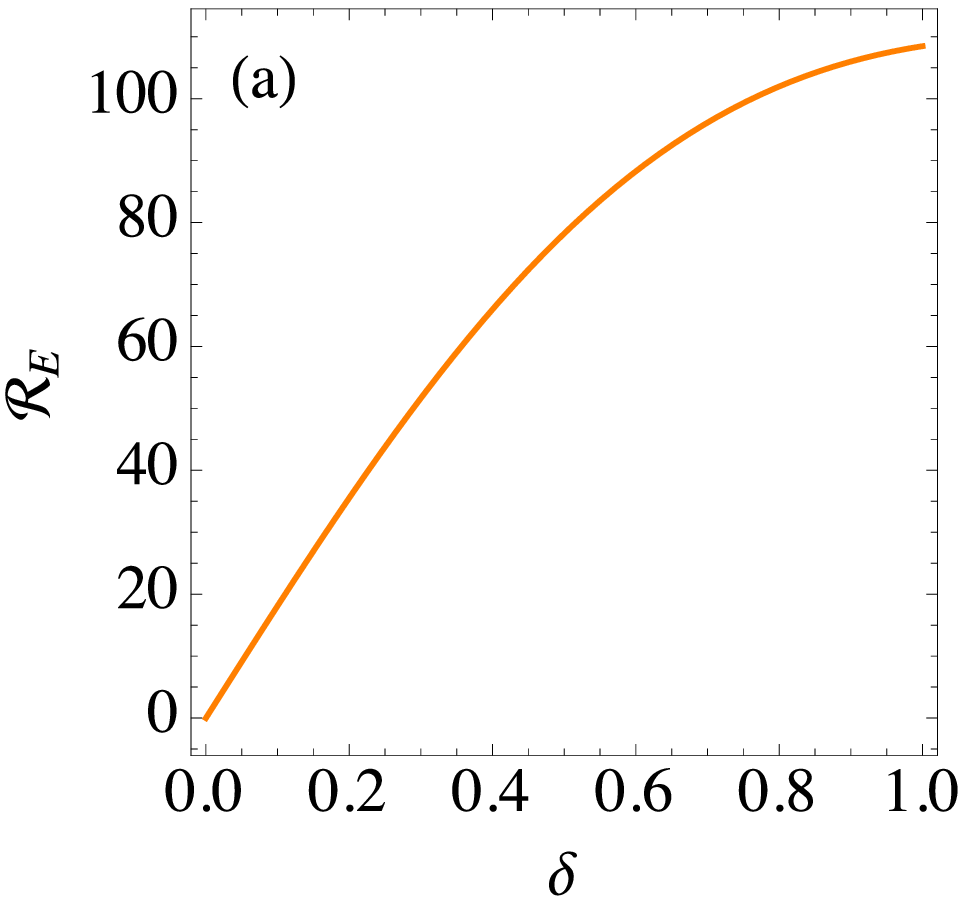}\quad
\includegraphics[width=0.40\textwidth]{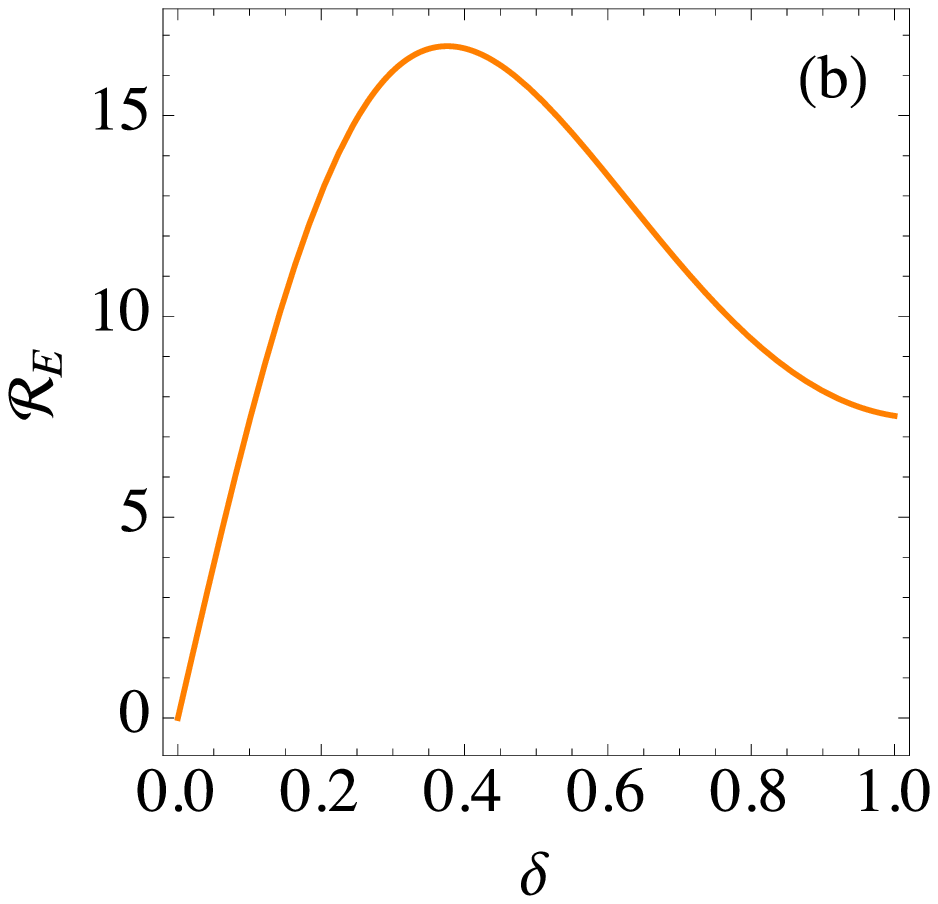}\\
\includegraphics[width=0.40\textwidth]{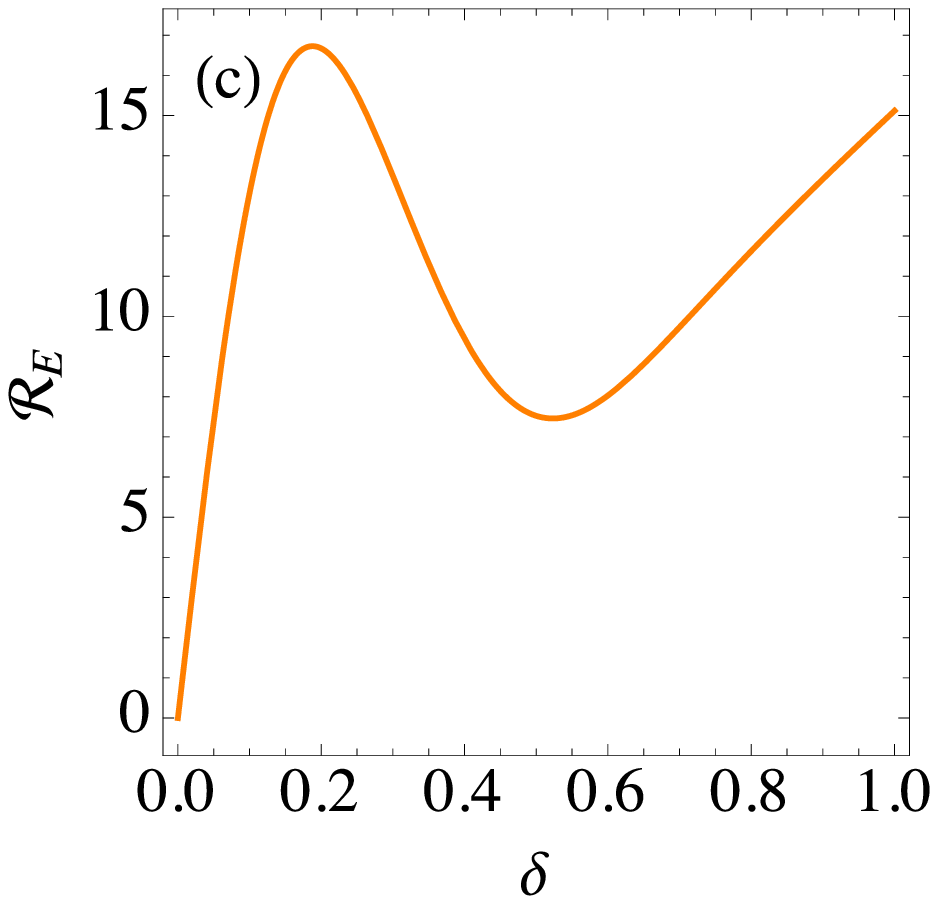}\quad
\includegraphics[width=0.40\textwidth]{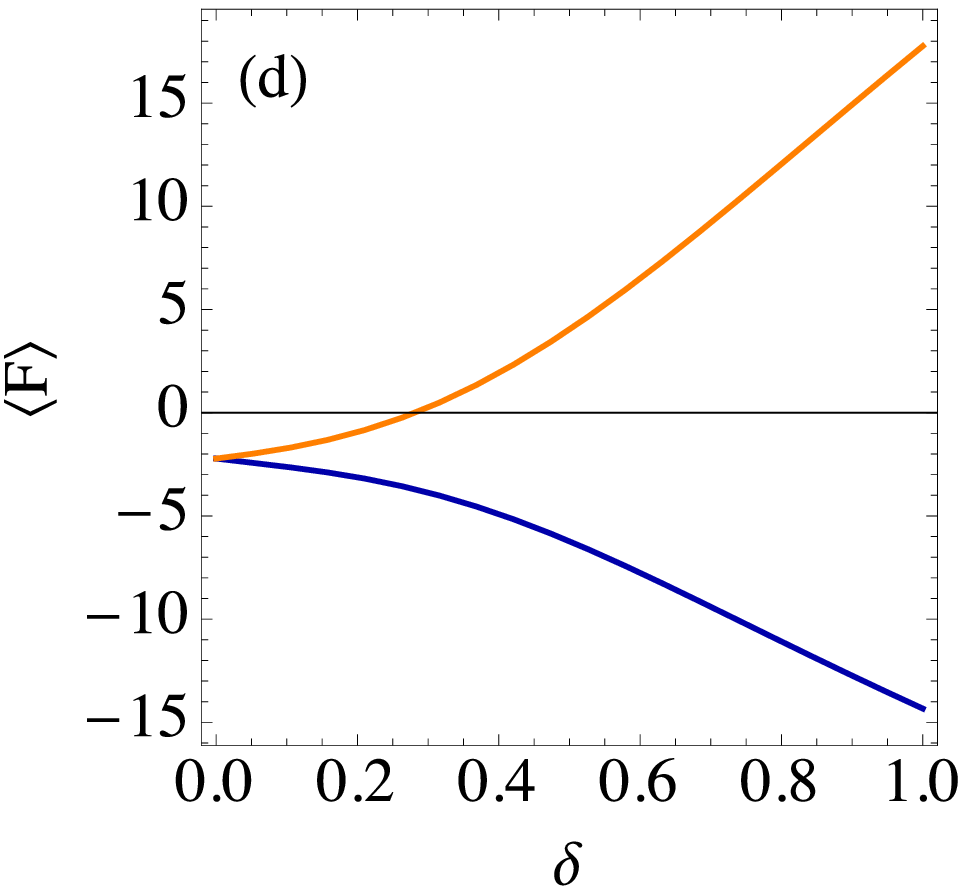}
\caption{(Color online) Energy rectification $\mathcal{R}_{E}$ versus parameter of asymmetry in several cases with $N=3$. a) Graded interaction between $S_{i}^{x}$ and $S_{i+1}^{x}$, $S_{i}^{y}$ and $S_{i+1}^{y}$:
$\alpha_{1,2} = 1 - \delta$,  $\alpha_{2,3} = 1 + \delta$; and homogeneous interaction between $S_{i}^{z}$ and $S_{i+1}^{z}$: $\Delta_{i,i+1} = \Delta = 1$. b) Graded XXX model:
 $\alpha_{1,2} = \Delta_{1,2} = 1 - \delta$,
$\alpha_{2,3} = \Delta_{2,3} = 1 + \delta$. c) Completely graded XXZ model: $\alpha_{1,2}  = 1 - \delta$, $\alpha_{2,3}  = 1 + \delta$,  $\Delta_{1,2} = 1 - \delta$, $\Delta_{2,3} = 1 + \delta$. In all these cases a), b) and c), we take $B = 0.1$, $f = \pm 0.1$. d) Energy current versus parameter of asymmetry in the model with graded interaction between $S_{i}^{z}$ and $S_{i+1}^{z}$, but
twisted XY boundary gradients:
$\langle S_{0}^{y} \rangle = \kappa =  \langle S_{N+1}^{x} \rangle$. Here,  $\Delta = 1$, $B=0.1$, $\alpha =1$, and $\kappa = 1/4$ for the upper (orange) curve; the lower (blue) curve follows for the system with inverted reservoirs, i.e.,
$\langle S_{0}^{x} \rangle = \kappa =  \langle S_{N+1}^{y} \rangle$.}
\end{figure}


We also investigate effects due to the introduction of a graded magnetic field $B$. The main new fact here is that rectification appears now also for the spin
current, as already observed in Ref.\cite{Landi} for the case of homogeneous XXZ (i.e., inner structures $\alpha$ and $\Delta$ constant along the chain). We observe that the energy current also rectifies in such case even with
inner homogeneous structures. Consistently, the rectification remains in both currents if we consider graded inner structures together with the graded magnetic field. As expected, there are many effects due to changes in the several parameters involved in these completely graded situations (e.g., regions in which the rectification increases or decreases, etc), but we do not further develop this investigation here.

A supplementary, interesting property present in some specific systems is the negative differential resistance (NDR),  the counterintuitive property of decreasing the energy current in the chain by increasing the pumping gradients at the edges. In the study of energy transport in classical anharmonic chains of oscillators, NDR is of fundamental importance in the building of thermal devices such as thermal transistors \cite{LiCasati}: it bears a
resemblance to the analogous property for the electric current in devices such as tunnel diodes. We have verified the existence of NDR in the quantum spin system investigated here: the quantum interaction is intricate enough to allow different
regimes of such a phenomenon. In Fig.5 we show some regions with the occurrence of NDR for the case of a chain with graded interaction between $S_{i}^{z}$ and $S_{i+1}^{z}$, and with the dissipator
targeting values of the $z$-spin at the boundaries. Specifically in Fig.5 a), the observed behavior mimics the one described in the classical graded chain of oscillators with
Fermi-Pasta-Ulam $\beta$ potentials (see Fig.4 b) in Ref.\cite{BLigraded}). In Fig.5 b) the behavior is similar to the case of a classical chain of oscillators with Frenkel-Kontorova interaction (see Fig.1 in Ref.\cite{BHuNDTR}).

\begin{figure}[h]
\includegraphics[width=0.40\textwidth]{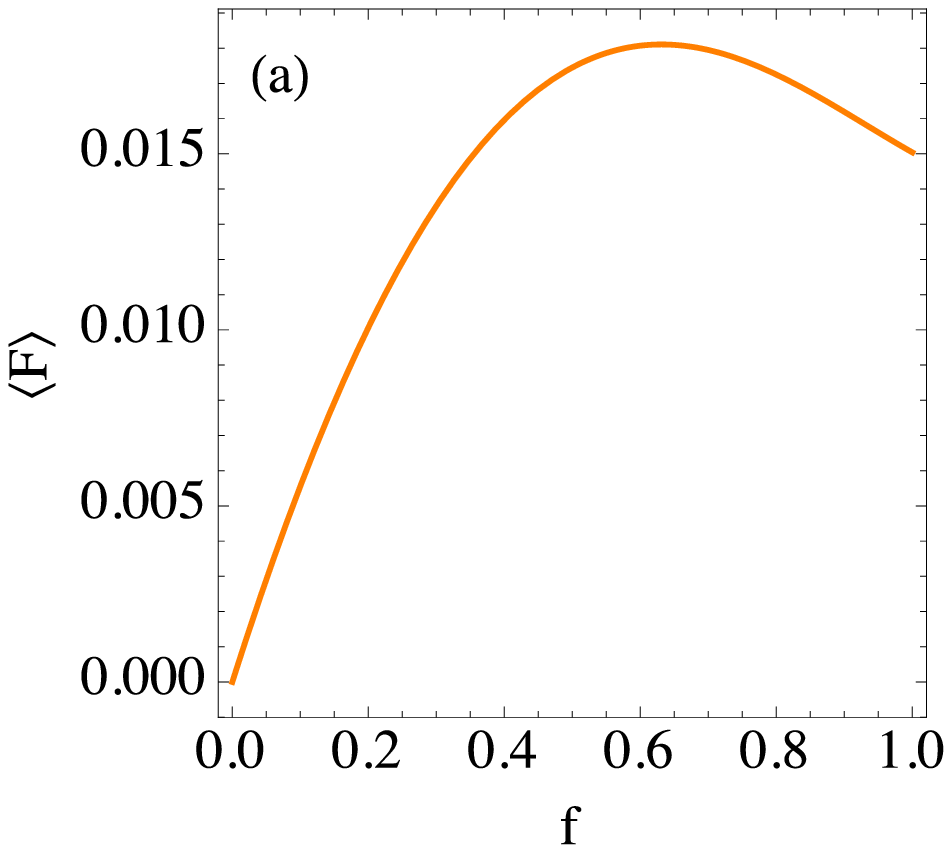}\quad
\includegraphics[width=0.40\textwidth]{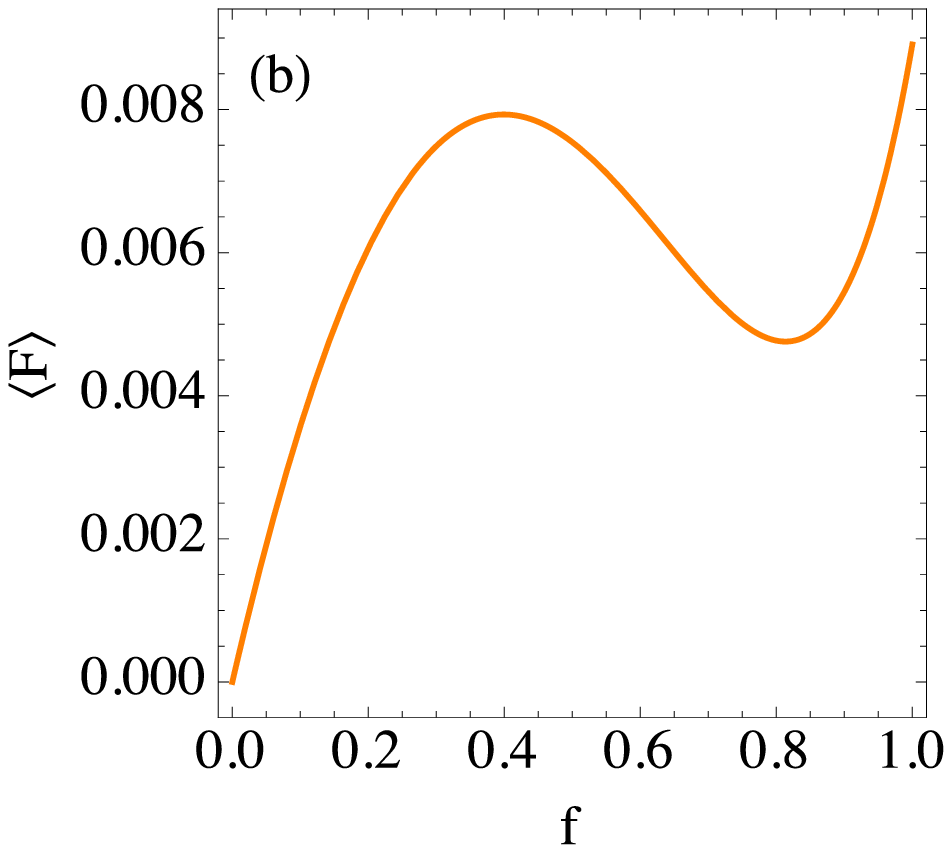}
\caption{NDR: energy current versus driving strength, for $N=3$ and graded interaction between $S_{i}^{z}$ and $S_{i+1}^{z}$. In a) $\alpha = 1$, $\Delta =1$, $\delta = 0.7$, $B = 0.1$. In b) $\alpha = 1$, $\Delta = 2$, $\delta = 0.8$, $B = 0.1$.}
\end{figure}

\section{Final Remarks}

In conclusion, by studying several types of graded XXZ models (and in different situations), we observe the generic occurrence of
energy rectification, even under a homogeneous  magnetic field $B$.
We recall that energy rectification is a ubiquitous phenomenon in classical systems \cite{WPC} represented by classical anharmonic chains of oscillators, and also
recall that these XXZ models, in a representation in terms of fermionic creation and annihilation operators, involve quartic (anharmonic) terms. Thus, in some
way, the present work extends the ubiquity of the energy rectification occurrence in graded anharmonic materials to  quantum spin systems.

Some final comments, regarding future investigations, seem appropriate.

First, we remark the importance of further studies concerning rectification properties
in larger systems in order to investigate, for example, possible decays with the system size. We also remark the interest of further analysis considering the energy transport in the spin chain submitted
to an energy imbalance at the boundaries (instead of the magnetization imbalance treated here).

Finally, we recall recent studies of classical anharmonic graded chains of oscillators with long range interactions, feature which significatively amplifies the rectification factor and avoids its decay with the system size \cite{CPC}, in order to announce the investigation of the graded
XXZ with long range interactions as a promising problem. Again, we stress that it is an issue of theoretical and experimental interest, since related models with long range interactions describe important real materials, as we may confirm with some well known examples:
 Coulomb crystals given by a trapped beryllium system, present candidates for many-qubit processors, may be engineered  to exhibit Ising-like
interactions with $J_{i,j} \propto 1/r^{\gamma}_{i,j}$, where $r_{i,j}$ is the distance between spin pairs and $0<\gamma<3$ \cite{Bri}; nanomagnets such as
permalloy may be lithographically manipulated to present different structures (e.g. spin ice), and their interactions typically decay as $1/r^3$, where $r$
denotes the distance between two nanodisks \cite{Wa}.

\vspace{0.5cm}

\section*{Appendix: Some Algebraic Expressions}

We list here some complete expressions obtained for the energy and spin currents, for $N=3$ and dissipators targeting the average values of the spin $S^{z}$ at the boundaries (denoted by $f$); see Sec.II.
For simplicity, we take the coupling strength $\gamma=1$. For the case of a homogeneous magnetic field $B$  and graded interaction $\Delta$, i.e., with the Hamiltonian
\begin{eqnarray*}
\mathcal{H} &=& \sum_{i=1}^{N-1}\left\{ \alpha\left( \sigma_{i}^{x}\sigma_{i+1}^{x} + \sigma_{i}^{y}\sigma_{i+1}^{y} \right) + \Delta_{i,i+1}\sigma_{i}^{z}\sigma_{i+1}^{z} \right\} \\
&& + ~B\sum_{i=1}^{N}\sigma_{i}^{z} ~,
\end{eqnarray*}
with $\Delta_{1,2} = \Delta - \delta$ and $\Delta_{2,3} = \Delta + \delta$, we obtain for the spin current
\begin{widetext}
\begin{eqnarray*}
\lefteqn{\langle J \rangle = 16f\alpha^{2} [589824\alpha^{8} + 32768\alpha^{6} (9 + 8(9 + 5f^{2} )\delta^{2}
  + 6\Delta^{2} ) (1 + 16\delta^{2} )^2 (256(3 - 2f^{2} )^{2}  \delta^{4} (9+(48 - 32f^{2} ) \Delta^{2} )^{2}} \\
 && - 32(-3 + 2f^{2} ) \delta^{2}  (9+16(-3+2f^{2} ) \Delta^{2}))+512\alpha^{4}  (99 + 256(15 - 6f^{2} + 2f^{4} )\delta^{4}
  - 48(-5 + 2f^{2} )\Delta^{2} \\
  && - 32\delta^{2} (-48( 1 + 2\Delta^{2})
  + f^2 (-7 + 24\Delta^{2}))) - 64\alpha^{2} (1 + 16\delta^{2})(256(-3 - f^{2} + 2f^{4} ) \delta^{4} + (6 + 4\Delta^{2})(-9 \\
  && + 16(-3 + 2f^{2} ) \Delta^{2}) - 16\delta^{2} (27 + 156\Delta^{2} + 32f^{4} \Delta^{2} - f^{2} (3 + 152^{2})))]/\\
 && (9437184\alpha^{10} + 65536\alpha^{8} (81 + 16(39 + 20f^{2} ) \delta^{2} + 48\Delta^{2} ) + 8192\alpha^{6} (135 + 256(21 - 3f^{2} + 2f^{4} ) \delta^{4} \\
 && - 48(-7 + 2f^{2} ) \Delta^{2} + 16\delta^{2} (126 + 272\Delta^{2} + f^{2} (21 - 48\Delta^{2} ) ) ) (1 + 16\delta^{2} )^{2} (-81 + 4096(-3 + 2f^{2} ) \delta^{6} \\
 && - 144(9 + 2f^{2} )\Delta^{2} + 256(-27 + 8f^{4} ) \Delta^{4} + 4096(-3 + 2f^{2} )\Delta^{6} + 256\delta^{4} (-27 + 8f^{4} + 48\Delta^{2} - 32f^{2} \Delta^{2}) \\
 && - 16\delta^{2} (81 + 288\Delta^{2} + 256f^{4}\Delta^{2} - 768\Delta^{4} + 2f^{2} (9 + 256\Delta^{4}))) - 512\alpha^{4} (-207  + 4096(-11 + f^{2} + 4f^{4} )\delta^{6} \\
 && + 48(-26 + 7f^{2} ) \Delta^{2} + 256(-3 + f^{2} ) \Delta^{4} - 256\delta^{4} (107 + 448\Delta^{2} + 8f^{4} (-1 + 8\Delta^{2}) - 8f^{2} (2 + 36\Delta^{2})) \\
 && - 16\delta^{2} (291 + 1664\Delta^{2} + 64f^{4}\Delta^{2} + 768\Delta^{4} - f^2 (7 + 592\Delta^{2} + 256\Delta^{4}))) + 16\alpha^{2} (1 + 16\delta^{2})(297 + 4096(13 \\
 && - 10f^{2} + 4f^{4} ) \delta^{6} - 96(-33 + 2f^{2} ) \Delta^{2} + 256(33 - 20f^{2} + 4f^{4} ) \Delta^{4} - 256\delta^{4} (-111 + 32\Delta^{2} + f^{2} (44 - 448\Delta^{2}) \\
 && + 4f^4 (-5 + 32\Delta^{2})) + 16\delta^{2} (315 + 2112\Delta^{2} + 6400\Delta^{4} + 128f^{4}\Delta^{2} (-3 + 8\Delta^{2}) - 2f^{2} (33 + 64\Delta^{2} + 2304\delta^{4}))))~.
\end{eqnarray*}
Expanding the expression in powers of the parameter of asymmetry $\delta$ up to order 2, we find
\begin{eqnarray*}
\lefteqn{\langle J \rangle = -\frac{(16(f\alpha^2 (-9 -768\alpha^4 + 16(-3 + 2f^2)\Delta^2 - 64\alpha^2 (3+4\Delta^2))))}{(9 +12288\alpha^6 +32(3 +2f^2)\Delta^2 + 256\Delta^4 +256\alpha^4 (15+16\Delta^2) +
\alpha^2 (336 -256(-7 +2f^2)\Delta^2))}}\\
&& + \left[256f\alpha^2 (-(3 + 48\alpha^2 + 16\Delta^2)^2 (65536\alpha^8  + 9(3 + 16\Delta^2) + 8192\alpha^6 (5+24\Delta^2) + 1024\alpha^4 (9 + 52\Delta^2)\right. \\
&& + 96\alpha^2 (9 + 40\Delta^2 + 128\Delta^4)) + 64f^4\Delta^2 (-81 + 184320\alpha^6 - 256\Delta^4 + 768\alpha^4 (27 + 160\Delta^2) \\
&& + 16\alpha^2(-45 + 192\Delta^2 + 1280\Delta^4)) + 2f^2 (9437184\alpha^{10} + 589824\alpha^8 (9 + 16\Delta^2) + 8192\alpha^6 (117 -408\Delta^2 + 2560\Delta^4)\\
&& + 1536\alpha^4 (27 - 456\Delta^2 - 1664\Delta^4 + 6144\Delta^6) + 3(-81 + 864\Delta^2 + 2304\Delta^4 + 8192\Delta^6)\\
&& + 16\alpha^2 (-243 -720\Delta^2 -10752\Delta^4 -28672\Delta^6 +65536\Delta^8)))/\\
&& ((9 + 192\alpha^2 + 768\alpha^4 + (48 -32f^2)\Delta^2)(9 +12288\alpha^6 +32(3 +2f^2)\Delta^2 + 256\Delta^4 +256\alpha^4 (15 + 16\Delta^2)\\
&& +\alpha^2 \left.(336 -256(-7 +2f^2)\Delta^2))^2)\right] \delta^2 ~.
\end{eqnarray*}

For the energy current, we have
\begin{eqnarray*}
\lefteqn{\langle F \rangle = 16f\alpha^{2}[2f\delta(196608\alpha^{6}\Delta^{2} - 256\alpha^{4} (9 + 256\delta^{4} - 192f^{2} \Delta^{2} - 256\Delta^{4} - 32\delta^{2} (-5 + 16(-3 + f^{2})\Delta^{2}))} \\
&& - 32\alpha^{2}(1 + 16\delta^{2})(256(2 + f^{2})\delta^{4} + 16\delta^{2}(21 + 9f^{2} + 104\Delta^{2}) - (9 + 16\Delta^{2})(-3 + 8(-1 + 2f^{2})\Delta^{2}))\\
&& + (1 + 16\delta^{2})^{2}(-81 + 256(-3 + 2f^{2})\delta^{4} - 768\Delta^{4} + 32f^{2} \Delta^{2}(-9 + 16\Delta^{2}) - 32\delta^{2}(18 - 48\Delta^{2} + f^{2}(-9 + 32\Delta^{2})))) \\
&& + {\bf B}(589824\alpha^{8} + 32768\alpha^{6} (9 + 8(9 + 5f^{2})\delta^{2} + 6\Delta^{2}) + (1 + 16\delta^{2})^{2} (256(3 - 2f^{2})^{2} \delta^{4} + (9 + (48 - 32f^{2})\Delta^{2})^{2} \\
&& - 32(-3 + 2f^{2})\delta^{2} (9 + 16(-3 + 2f^{2})\Delta^{2})) + 512\alpha^{4}(99 + 256(15 - 6f^{2} + 2f^{4})\delta^{4} - 48(-5 + 2f^{2})\Delta^{2} \\
&& - 32\delta^{2}(-48(1 + 2\Delta^{2}) + f^{2}(-7 + 24\Delta^{2}))) - 64\alpha^{2}(1 + 16\delta^{2})(256(-3 - f^{2} + 2f^{4})\delta^{4} + (6 + 4\Delta^{2})(-9  \\
&& + 16(-3 + 2f^{2})\Delta^{2}) - 16\delta^{2}(27 + 156\Delta^{2}  + 32f^{4}\Delta^{2} - f^{2}(3 + 152\Delta^{2}))))]/\\
&& (9437184\alpha^{10} + 65536\alpha^{8}(81 + 16(39 + 20f^{2})\delta^{2} + 48\Delta^{2}) + 8192\alpha^{6}(135 + 256(21 - 3f^{2} + 2f^{4})\delta^{4} - 48(-7 + 2f^{2})\Delta^{2} \\
&& + 16\delta^{2}(126 + 272\Delta^{2} + f^{2}(21 - 48\Delta^{2}))) - (1 + 16\delta^{2})^{2}(-81 + 4096(-3 + 2f^{2})\delta^{6} - 144(9 + 2f^{2})\Delta^{2} \\
&& + 256(-27 + 8f^{4})\Delta^{4} + 4096(-3 + 2f^{2})\Delta^{6} + 256\delta^{4}(-27 + 8f^{4} + 48\Delta^{2} - 32f^{2}\Delta^{2}) - 16\delta^{2}(81 + 288\Delta^{2} \\
&& + 256f^{4}\Delta^{2} - 768\Delta^{4} + 2f^{2}(9 + 256\Delta^{4}))) - 512\alpha^{4}(-207 + 4096(-11 + f^{2} + 4f^{4})\delta^{6} + 48(-26 + 7f^{2})\Delta^{2} \\
&& + 256(-3 + f^{2})\Delta^{4} - 256\delta^{4}(107 + 448\Delta^{2} + 8f^{4} (-1 + 8\Delta^{2}) - 8f^{2}(2 + 36\Delta^{2})) - 16\delta^{2}(291 + 1664\Delta^{2} \\
&& + 64f^{4}\Delta^{2} + 768\Delta^{4} - f^{2}(7 + 592\Delta^{2} + 256\Delta^{4}))) + 16\alpha^{2}(1 + 16\delta^{2})(297 + 4096(13 - 10f^{2} + 4f^{4})\delta^{6} \\
&& - 96(-33 + 2f^{2})\Delta^{2} + 256(33 - 20f^{2} + 4f^{4})\Delta^{4} - 256\delta^{4}(-111 + 32\Delta^{2} + f^{2}(44 - 448\Delta^{2}) + 4f^{4}(-5 + 32\Delta^{2})) \\
&& + 16\delta^{2}(315 + 2112\Delta^{2} + 6400\Delta^{4} + 128f^{4}\Delta^{2}(-3 + 8\Delta^{2})-2f^{2}( 33 + 64\Delta^{2} + 2304\Delta^{4})))) ~,
\end{eqnarray*}
which, in powers of $\delta$, up to $\delta^{2}$,
\begin{eqnarray*}
\lefteqn{\langle F \rangle = \frac{-(16(Bf\alpha^2 (-9 -768\alpha^4 + 16(-3 + 2f^2)\Delta^2 - 64\alpha^2 (3 + 4\Delta^2))))}{(9 + 12288\alpha^6 + 32(3 + 2f^2)\Delta^2 + 256\Delta^4 + 256\alpha^4 (15 + 16\Delta^2)
+ \alpha^2 (336 - 256(-7 + 2f^2)\Delta^2))}}\\
&& + \left[32f^2 \alpha^2 (-81 + 196608\alpha^6 \Delta^2 - 768\Delta^4 + 32f^2 \Delta^2 (-9 + 16\Delta^2) + 32\alpha^2 (9 + 16\Delta^2)(-3 + 8(-1 + 2f^2)\Delta^2)\right. \\
&& + 256\alpha^4 (-9 + 192f^2 \Delta^2 + 256\Delta^4))/(81 + 9437184\alpha^{10} + 144(9 + 2f^2)\Delta^2 - 256(-27 + 8f^4)\Delta^4 - 4096(-3 + 2f^2)\Delta^6 \\
&& + 196608\alpha^8 (27 + 16\Delta^2) - 24576\alpha^6 (-45 + 16(-7 + 2f^2)\Delta^2) - 512\alpha^4 (-207 + 48(-26 + 7f^2)\Delta^2 + 256(-3 + f^2)\Delta^4) \\
&& \left. + 16\alpha^2 (297 - 96(-33 +2f^2)\Delta^2 + 256(33 -20f^2 + 4f^4)\Delta^4)) \right]\delta\\
&& + \left[256Bf\alpha^2 (-(3 + 48\alpha^2 + 16\Delta^2)^2 (65536\alpha^8 + 9(3 + 16\Delta^2) + 8192\alpha^6 (5 + 24\Delta^2)\right. \\
&& + 1024\alpha^4 (9 + 52\Delta^2) + 96\alpha^2 (9 + 40\Delta^2 + 128\Delta^4)) + 64f^4 \Delta^2 (-81 + 184320\alpha^6 - 256\Delta^4 + 768\alpha^4 (27 + 160\Delta^2) \\
&& + 16\alpha^2 (-45 + 192\Delta^2 + 1280\Delta^4)) + 2f^2 (9437184\alpha^{10} + 589824\alpha^8 (9 + 16\Delta^2) + 8192\alpha^6 (117 - 408\Delta^2 + 2560\Delta^4) \\
&& + 1536\alpha^4 (27 - 456\Delta^2 - 1664\Delta^4 + 6144\Delta^6) + 3(-81 + 864\Delta^2 + 2304\Delta^4 + 8192\Delta^6) + 16\alpha^2 (-243 -720\Delta^2 -10752\Delta^4 \\
&& -28672\Delta^6 +65536\Delta^8)))/((9 + 192\alpha^2 + 768\alpha^4 + (48 -32f^2)\Delta^2)(9 + 12288\alpha^6 + 32(3 + 2f^2)\Delta^2 + 256\Delta^4 \\
&&\left.  + 256\alpha^4 (15 + 16\Delta^2) + \alpha^2 (336 - 256(-7 + 2f^2)\Delta^2))^2)\right] \delta^2 ~.
\end{eqnarray*}

The expression and the behavior of $\langle F \rangle$ becomes more transparent if we also take the expansion in powers of the driving strength $f$, i.e., for small $f$ and $\delta$ we obtain
\begin{eqnarray*}
\lefteqn{\langle F \rangle =   fB \frac{48 \left(16
   \alpha ^2+3\right) \alpha ^2 }{768 \alpha ^4+192 \alpha ^2+48 \Delta ^2+9}}\\
   &&
+   f^2 \delta \frac{32 \alpha ^2  \left(196608 \alpha ^6 \Delta ^2+65536 \alpha ^4 \Delta ^4-2304 \alpha ^4-4096 \alpha ^2 \Delta ^4-3840 \alpha ^2 \Delta ^2-864
   \alpha ^2-768 \Delta ^4-81\right)}{3 \left(48 \alpha ^2+16 \Delta ^2+3\right) \left(256 \alpha ^4+64 \alpha ^2+16 \Delta ^2+3\right)^2}~.
\end{eqnarray*}
\end{widetext}

A superficial analysis of the formulas is enough to make transparent some claimed results: the structure of the energy current $\langle F \rangle $ given as $\langle F \rangle = \langle F^{XXZ} \rangle + B \langle J \rangle$;
the nonvanishing of $\langle F^{XXZ} \rangle$ in the graded case for $B=0$, and the preferential direction of $\langle F \rangle$ in such case (it is an even function of $f$, and so, it does not change with a change in the sign
of $f$); and, finally, the absence of rectification for the spin current $\langle J \rangle$ and its occurrence for the energy current $\langle F \rangle$. Note also that this rectification for the energy current $\langle F \rangle$ vanishes in the XX model, i.e., as $\Delta$ and $\delta$ go to zero.

\vspace*{1 cm} {\bf Acknowledgments:} This work was partially supported by CNPq (Brazil). G. T. Landi acknowledges the financial support from the S\~ao Paulo Research Foundation (FAPESP), under project grant 2014/01218-2.

\end{document}